\documentclass[a4paper,fleqn]{cas-dc}

\usepackage[numbers]{natbib}
\usepackage{amsmath,amssymb,amsfonts}
\usepackage{algorithmic}
\usepackage[ruled]{algorithm2e}
\usepackage{graphicx,subfig}
\usepackage{epsfig}
\usepackage{caption}

\usepackage{color}

\begin{document}

\shortauthors{Yang Li et~al.}
%\shorttitle{Stochastic Blockmodeling and Fast Learning for Large-scale Signed Networks}
\shorttitle{SSBM: A Signed Stochastic Block Model for Multiple Structure Discovery in Large-Scale Exploratory Signed Networks}

%\title [mode = title]{Stochastic Blockmodeling and Fast Learning for Large-scale Signed Networks}
\title [mode = title]{SSBM: A Signed Stochastic Block Model for Multiple Structure Discovery in Large-Scale Exploratory Signed Networks}
%\title [mode = title]{SSBM: A Singed Stochastic Blockmodel for Large-scale Network Analysis}

\author[3]{Yang Li}
\author[1,2]{Bo Yang}[orcid=0000-0003-1927-8419] \corref{cor1}
\author[4]{Xuehua Zhao}
\author[1,5]{Zhejian Yang}
\author[1,5]{Hechang Chen}[orcid=0000-0001-7835-9556] \corref{cor1}

\address[1]{Key Laboratory of Symbolic Computation and Knowledge Engineering, Ministry of Education, Changchun, 130012, China}
\address[2]{College of Computer Science and Technology, Jilin University, Changchun, 130012, China}
\address[3]{Aviation University of Air Force, Changchun, 130062, China}
\address[4]{School of Digital Media, Shenzhen Institute of Information Technology, Shenzhen, 518055, China}
\address[5]{School of Artificial Intelligence, Jilin University, Changchun, 130012, China}
\cortext[cor1]{Corresponding authors: Bo Yang and Hechang Chen}

%0000-0001-7835-9556

\begin{abstract}
Signed network structure discovery has received extensive attention and has become a research focus in the field of network science. However, most of the existing studies are focused on the networks with a single structure, e.g., community or bipartite, while ignoring multiple structures, e.g., the coexistence of community and bipartite structures. Furthermore, existing studies were faced with challenge regarding large-scale signed networks due to their high time complexity, especially when determining the number of clusters in the observed network without any prior knowledge. In view of this, we propose a mathematically principled method for signed network multiple structure discovery named the Signed Stochastic Block Model (SSBM). The SSBM can capture the multiple structures contained in signed networks, e.g., community, bipartite, and coexistence of them, by adopting a probabilistic model. Moreover, by integrating the minimum message length (MML) criterion and component-wise EM (CEM) algorithm, a scalable learning algorithm that has the ability of model selection is proposed to handle large-scale signed networks. By comparing state-of-the-art methods on synthetic and real-world signed networks, extensive experimental results demonstrate the effectiveness and efficiency of SSBM in discovering large-scale exploratory signed networks with multiple structures.
\end{abstract}


\maketitle

\section{Introduction}

Structure discovery for signed networks with positive and negative edges has received extensive attention and has become a research focus in the field of network science \cite{Chen_and_Wang,Wang_and_Gong,Zhao_and_Yang,Zhao_and_Liu}.
Different from unsigned networks containing only one kind of edge describing a homogenous relationship  \cite{Liu_and_Yang,He_and_Zhang,Rahimi_and_Abdollahpouri},
the positive edges in signed networks usually represent trust, like, or support relationships, while the negative edges usually represent distrust, dislike, or oppose relationships \cite{Lui_and_Su,Tang_and_Chang}.
Therefore, signed networks can characterize different relationships between individuals by adding positive and negative signs.

In recent years, researchers have found that the community, i.e., a dense subnetwork, is one of the most common structures in real-world networks \cite{Newman}.
Subsequently, some community discovery methods have been proposed and can be classified into two categories: discriminant and principled methods.
For discriminant methods, optimization objectives, e.g., modularity, or heuristics, e.g., random walk model, should be predefined according to specific network characteristics \cite{Doreian_and_Mrvar,Bansal_and_Blum,Traag_and_Bruggeman,Yang_and_Cheung,Anchuri_and_Magdon,Li_and_He,Zhu_and_Ma}.
However, discriminant methods are inflexible in practical applications due to the difficulty of designing appropriate objective functions. 
The principled methods are usually used to detect structures in signed networks because probability models can capture the intrinsic features of different structures when fitting the observed networks  \cite{Zhao_and_Yang,Yang_and_Liu}.

Despite the success that previous studies achieved in structure discovery, there are two challenges remain unsolved: 
1) Generalization: both discriminant methods and principled methods can mostly discover one single structure, e.g., community or bipartite, but cannot detect multiple structures, e.g., the coexistence of community and bipartite.
2) Scalability: most principled methods have high time complexity due to parameter estimation and model selection, i.e., determining the number of clusters $K$. 
To select an optimal model, these methods have to traverse all possible $K$ values and then calculate the parameters of all the possible models \cite{Zhao_and_Liu,Yang_and_Liu}, leading to high time complexity for a slightly large-scale network.

In view of this, a generalized model for characterizing multiple structures and a scalable learning algorithm for large-scale exploratory signed networks are proposed in this paper.  
The contributions are summarized as follows:

    1) \emph{A new reparameterized signed stochastic block model, namely SSBM, is proposed to characterize the multiple structures in the signed networks}. 
    
In the SSBM, a new parameter $\Lambda$ is introduced to reparameterize the parameter $\Pi$ in the standard SBM. 
The reparameterized model continues the idea of using block structure to characterize network structure, and the property of structural equivalence in the standard SBM, i.e., the nodes in the same block have similar connection patterns. 
This allows SSBM to characterize a single structure, e.g., community or bipartite, and even more complex multiple structures, e.g., their coexistence in the signed networks. 
In addition, the reparameterization can fundamentally solve the high time complexity problem encountered by most existing SBM learning methods, e.g.,  $O(K^2n^2)$ when $K$ is known; otherwise, $O(n^5)$. This makes it possible to detect and analyze multiple structures of large-scale signed networks.

    2) \emph{A scalable learning algorithm SSBM with model selection ability is proposed for large-scale exploratory signed networks.} 

Model selection ability indicates whether an algorithm can discover network structures without prior knowledge. 
For most existing signed network structure discovery methods, a serial learning mechanism is usually employed to perform parameter estimation and model selection alternately in the model space. 
The time complexity resulting from this mechanism is generally $O(n^5)$. 
For this purpose, a scalable learning algorithm SSBM by integrating the minimum message length (MML) \cite{Wallace_and_Boulton} and component-wise EM (CEM) algorithm \cite{Celeus_and_Chrtien} is proposed. 
The SSBM can synchronously perform parameter estimation and model selection in the block space $[K_{min}, K_{max}]$, which can effectively reduce the learning time from $O(n^5)$ to $O(n^3)$. 
This parallel learning mechanism is crucial for realizing multiple structure discovery of large-scale exploratory signed networks. 

    3) \emph{Experimental results on synthetic and real-world signed networks in terms of generalization, robustness, and scalability demonstrate superiority of SSBM}.

For generalization, the experimental results on various networks validate the effectiveness of SSBM in discovering multiple structures. 
For robustness, according to the learned parameters $\Lambda$, the densities of different noises, i.e., negative edges within a block and positive edges between blocks, can be denoted explicitly. 
Therefore, the SSBM can overcome the influence of different noise types and densities and accurately discover the multiple structures in the signed networks. 
For scalability, the experimental results on large-scale synthetic and real-world signed networks confirm that SSBM can effectively handle networks with tens of thousands of nodes in minute-level time. 
This significant advantage allows SSBM to discover multiple structures of large-scale exploratory signed networks.

The rest of the paper is organized as follows. The proposed model and learning algorithm are described in detail in Section 2. The experimental results to validate the effectiveness and efficiency of SSBM are presented in Section 3.
Section 4 summarizes the related works in terms of discriminant and principled methods, and we conclude the paper in Section 5.

\section{Methodology}
In this section, a novel signed stochastic block model, namely SSBM, is introduced by reparameterizing the standard SBM to discover multiple structures in signed networks. Then, a scalable learning algorithm is proposed to perform parameter estimation and model selection simultaneously.

\subsection{Signed Stochastic Block Model}
%\hechang{Definition of traditional SBM should be given.}\liyang{Resolved.}

The standard SBM that only discovers structures contained in unsigned networks can be formalized as \cite{Snijders_and_Nowicki}: 
\begin{equation}
\label{EQ0}
X = (K,Z,\Phi,\Pi)
\end{equation}
where $K$ denotes the number of blocks in a network consisting of $n$ nodes. The latent variable $Z$ is a $n \times K$ dimensional matrix, and $z_{ik}=1$ if node $i$ is allocated to block $k$; otherwise, $z_{ik} = 0$.
$\Phi=(\phi_1, \cdots, \phi_K)$ is a $K$ dimensional vector, and $\phi_k$ represents the probability of allocating a node to block $k$. $\Pi$ is a $K \times K$ dimensional matrix, and the element $\pi_{kq}$ represents the probability of generating an edge between the nodes in  block $k$ and block $q$.

According to the above definition, the standard SBM only can characterize unsigned network structure due to matrix $\Pi$.
For this problem, a new parameter $\Lambda$ is introduced to reparameterize $\Pi$, and $\Lambda$ represents the probability matrix that positive, negative, or null edges will be generated from a block to a node. Then a new signed stochastic block model is presented to characterize multiple structures in the signed networks at a finer granularity.

\begin{figure}[htbp]
  \centering
  \includegraphics[width=0.3 \textwidth]{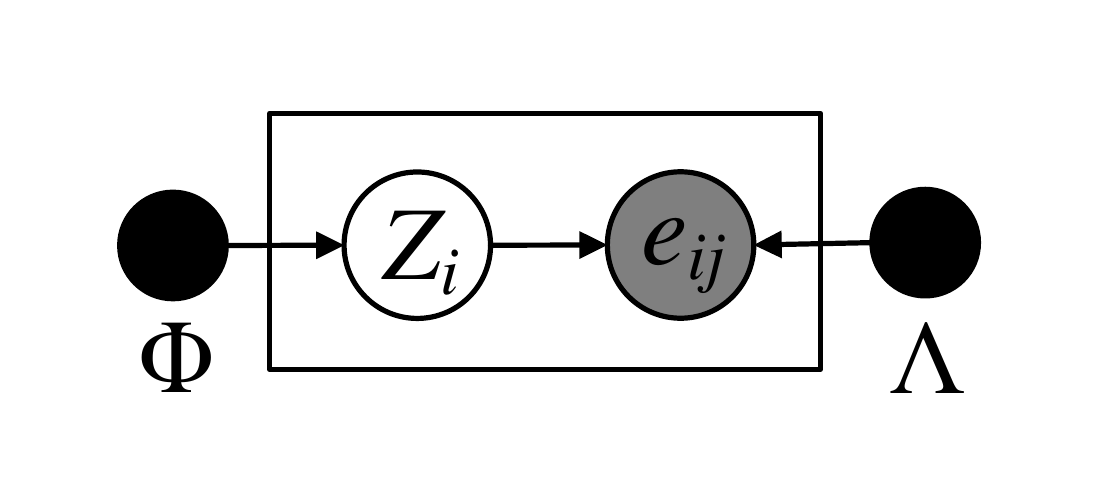}
  \caption{\normalsize{Graphical model of SSBM.}}\label{GraphicalSSBM}
\end{figure}

Assuming that $E_{n \times n}$ is an adjacent matrix of an observed signed network $N$ containing $n$ nodes. The element $e_{ij}$ equals to 1, -1, or 0, denoting a positive, negative, or null edge existing between node $i$ and node $j$.
Then, a new signed stochastic block model, named SSBM, can be formulated as follows:
\begin{equation}
\label{EQ1}
X = (K,Z,\Phi,\Lambda)
\end{equation}
where the parameters $K, Z, \Phi$ are identical to them in the standard SBM, i.e., $K$ denotes the number of blocks, $Z$ is a $n \times K$ dimensional latent variable indicating node assignments, and $\Phi$ is a $K$ dimensional vector representing probability distribution of allocating a node to different blocks. $\Lambda$ is a $K \times n \times 3$ dimensional block-to-node connection matrix, and the element $\lambda_{kj}=(\lambda_{kj1}, \lambda_{kj2}, \lambda_{kj3})$ represents probability generating a positive, negative, or null edge from a node in block $k$ and node $j$, respectively.
 Fig.\ref{GraphicalSSBM} shows a graphical model of SSBM.
%\hechang{The SSBM model can also be seen as a generative model, by with xxx.}

\begin{figure*}[htbp]
  \centering
    \includegraphics[width=0.9 \textwidth]{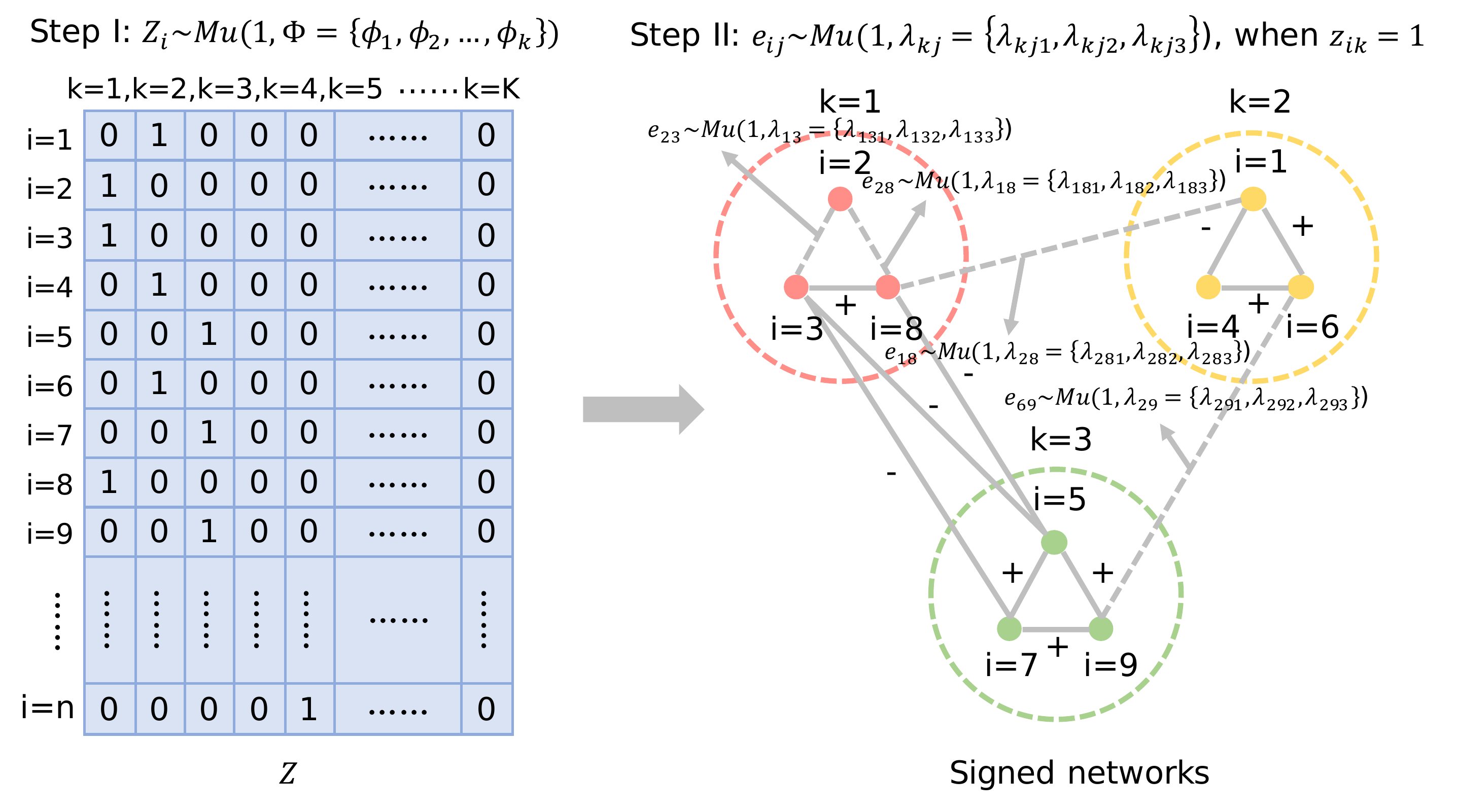}
  \captionsetup{justification=centering}
  \caption{\normalsize{The steps of SSBM generating  signed networks.The solid lines between two nodes denote the generated positive or negative edges labeled as "+" or "-", respectively. The dotted lines denote edges to be generated, and the signs of edges are determined by multinomial distribution with the parameter $\lambda_{kj}$ formalized as $Mu(1,\lambda_{kj}=\{\lambda_{kj1},\lambda_{kj2},\lambda_{kj3}\})$.}}\label{generativeprocess}
  \label{synthetic networks}
\end{figure*}

Note that SSBM is also a generative model that can generate various signed networks with block structures. A signed network can be generated by the following steps:

Step I: Allocate a node $i$ to a block $k$ according to the probability $\phi_k$;

Step II: If node $i$ is in block $k$, then generating a positive, negative, or null edge between node $i$ and node $j$ according to multinomial distribution $\lambda_{kj}=(\lambda_{kj1}, \lambda_{kj2}, \lambda_{kj3})$.

The graphical model of SSBM and the steps of SSBM generating signed networks as mentioned above are  exhibited in Fig.\ref{GraphicalSSBM} and  Fig.\ref{generativeprocess}, respectively. It is clear that $e_{ij}$ is dependent on $Z_{i}$ and $\Lambda$, and $Z_{i}$ is only dependent on $\Phi$. Therefore, the likelihood of an observed signed network $N$ and the latent variable $Z$, namely, the complete data likelihood $p(N,Z|K,\Phi,\Lambda)$, can be formalized as follows:
\begin{equation}
\begin{split}
\label{EQ4_1}
p(N,Z|K,\Phi,\Lambda) = p(Z|K,\Phi)p(N|K,Z,\Lambda)
\end{split}
\end{equation}
For Equation (\ref{EQ4_1}),
\begin{equation}
\begin{split}
\label{EQ4_2}
p(Z|K,\Phi) = \prod_{i=1}^{n}\prod_{k=1}^{K}\phi_{k}^{z_{ik}}
\end{split}
\end{equation}
\begin{equation}
\begin{split}
\label{EQ4_3}
p(N|K,Z,\Lambda) = \prod_{i=1}^{n}\prod_{k=1}^{K}\prod_{j=1}^{n}\prod_{h=1}^{3}\lambda_{kjh}^{z_{ik}\delta(a_{ij},2 - h)}
\end{split}
\end{equation}
%
%\hechang{How to interpret the complete data likelihood?}\liyang{Resolved.}
where $h \in \{1, 2, 3\}$, and when $x = y$, $\delta(x,y)=1$; otherwise, $\delta(x,y)=0$. Then the log form of Equation (\ref{EQ4_1}), i.e., the complete data log likelihood is:
\begin{equation}
\begin{split}
\label{EQ5}
\log p(N,Z|K,\Phi,\Lambda) = \sum_{i=1}^{n}\sum_{k=1}^{K}z_{ik}\log \phi_k\\
+ \sum_{i=1}^{n}\sum_{k=1}^{K}\sum_{j=1}^{n}\sum_{h=1}^{3}z_{ik}\delta(a_{ij},2 - h)\log \lambda_{kjh}
\end{split}
\end{equation}
%
%\hechang{Add some comments for equation 2 and 3 at the end of 2.1.}
%\liyang{Resolved.}

\subsection{Scalable Learning Algorithm}

By combining MML criterion with CEM algorithm, a scalable learning algorithm is proposed to carry out parameter estimation and model selection simultaneously.
The CEM algorithm is used to learn the parameters of blocks, and MML is used to evaluate and select blocks.
If one block is of poor quality, e.g., empty, then it is discarded and no longer computed in the subsequent iterations until convergence.
Different from most of the existing methods, our proposed learning algorithm focuses on  block space rather than model space.
The time complexity can be reduced when selecting models, and the efficiency can be improved significantly.

\subsubsection{Cost Function of SSBM}
%The cost function of the SSBM can be derived based on MML [1][3]:
The detailed deduction of SSBM cost function based on MML criterion is presented in this section.
Specificially, the cost function of MML is \cite{Lanterman A D, M.A.T. Figueiredo}:
\begin{equation}
\begin{split}
\label{loss function}
C(N,g) =& - log p(N|g) - \log p(g) + {\frac{1}{2}}\log |\textbf{I}(g)|  \\
&+ {\frac{d}{2}}(1 + log \kappa_d)
\end{split}
\end{equation}
where $N$ represents the observed signed network, and $g$ denotes model parameters, i.e., $(K, \Phi, \Lambda)$ in SSBM.
$d$ is the dimension of $g$, $\kappa_d \approx (2\pi e)^{-1}$ when $d$ is large.
$\textbf{I}(g)\equiv -\mathbb{E}_{p(N|g)}[D_{g}^{2}\log p(N|g)]$ is the Fisher information matrix, $\mathbb{E}$ and $|\textbf{I}(g)|$
denote expectation and determinant, respectively.

Since $\textbf{I}(g)$ cannot be calculated analytically, an upper bound of $\textbf{I}(g)$ is constructed by the Fisher information matrix of complete data likelihood:
\begin{equation}
\label{Icg}
\textbf{I}_c(g)\equiv -\mathbb{E}_{p(N,Z|g)}[D_{g}^{2}\log p(N,Z|g)]
\end{equation}
where $D_{g}^{2}$ is the second derivative on $g$.
The $C(N,g)$ can be optimized by minimizing $\textbf{I}_c(g)$ \cite{D.M. Titterington}.
%
%\begin{equation}
%\begin{split}
%\label{fisher matrix}
%\textbf{I}_{c}(g)&={block-diag}\{n\omega_{1}^{-1},...,n\omega_{K}^{-1},\\
%&n\omega_{1}\theta_{111}^{-1},...,n\omega_{1}\theta_{113}^{-1},\\
%&\ldots \ldots,\\
%&n\omega_{1}\theta_{1n1}^{-1},...,n\omega_{1}\theta_{1n3}^{-1},\\
%&\ldots \ldots,\\
%&n\omega_{K}\theta_{K11}^{-1},...,n\omega_{K}\theta_{K13}^{-1},\\
%&\ldots \ldots,\\
%&n\omega_{K}\theta_{Kn1}^{-1},...,n\omega_{K}\theta_{Kn3}^{-1}\}
%\end{split}
%\end{equation}

We let $g=(K, \Phi, \Lambda)$, $\textbf{I}_c(g)$, a $3Kn+K$ dimensional diagonal matrix where diagonal elements are the second partial derivatives of log$p(N,Z|g)$.
According to Equation (\ref{Icg}), we have:
\begin{equation}
\label{phi}
-\mathbb{E}_{p(N,Z|g)}[\frac{\partial^2 \log p(N,Z|g)}{\partial {\phi_{k}}^2}] =n{\phi_{k}}^{-1}
\end{equation}

\begin{equation}
\label{lambda}
-\mathbb{E}_{p(N,Z|g)}[\frac{\partial^2 \log p(N,Z|g)}{\partial {\lambda_{kjh}}^2}] = n\phi_{k}{\lambda_{kjh}}^{-1}
\end{equation}

Then the determinant of the $3Kn+K$ dimensional diagonal matrix $\textbf{I}_c(g)$ is as follows:
\begin{equation}
\label{determinant}
|\textbf{I}_c(g)| = n^{3nK+K}\prod_{k=1}^{K}\phi_{k}^{-1}\prod_{q=1}^{K}\prod_{j=1}^{n}\prod_{h=1}^{3}\phi_{q}
\lambda_{qjh}^{-1}
\end{equation}

A noninformative prior is used to formally characterize $p(g)$ due to $\Phi$ independent on $\Lambda$.
\begin{equation}
\label{g prior}
p(g) = p(\phi_{1},...,\phi_{k})\prod_{k=1}^{K}\prod_{j=1}^{n}p(\lambda_{kj1},\lambda_{kj2},\lambda_{kj3})
\end{equation}

The standard Jeffrey prior \cite{Murphy} is adopted to characterize  $p(\phi_k, \cdots, \phi_K)$ and
$p(\lambda_{kj1}, \lambda_{kj2}, \lambda_{kj3})$.
\begin{equation}
\label{omega prior}
p(\phi_{1},...,\phi_{k}) \propto (\prod_{k=1}^{K}\phi_{k})^{-\frac{1}{2}}
\end{equation}
\begin{equation}
\label{theta prior}
p(\lambda_{kj1},\lambda_{kj2},\lambda_{kj3}) \propto (\lambda_{kj1}\lambda_{kj2}\lambda_{kj3})^{-\frac{1}{2}}
\end{equation}

According to Equations (\ref{g prior}), (\ref{omega prior}), and (\ref{theta prior}), -$\log p(g)$ in Equation (\ref{loss function}) can be formalized as follows:
\begin{equation}
\label{EQ20}
-\log p(g) = -\frac{1}{2}\sum_{k=1}^{K}\log \phi_{k}^{-1} - \frac{1}{2}\sum_{k=1}^{K}\sum_{j=1}^{n}\sum_{h=1}^{3}\log \lambda_{kjh}^{-1}
\end{equation}

Finally, the cost function of SSBM is:
\begin{equation}
\begin{split}
\label{SSBM loss function}
C(N,g)=-\log p(N|g) + \frac{K(c + 1)}{2}\log n \\
+ \frac{c}{2}\sum_{k = 1}^{K}\log \phi_{k} + \frac{d}{2}(1 + \log \kappa_d)
\end{split}
\end{equation}
where $c$ is the number of parameters in a single block, and $g$ denotes the model parameters, i.e., $(K, \Phi, \Lambda)$.

From the perspective of information coding, Equation (\ref{SSBM loss function}) is essentially the sum of data coding lengths, i.e., $-\log p(N|g)$, and model coding length, i.e., $\frac{K(c + 1)}{2}\log n + \frac{c}{2}\sum_{k = 1}^{K}\log \phi_{k} + \frac{d}{2}(1 + \log \kappa_d)$.
Optimizing the cost function is to minimize information coding length.
Because the parameters of an empty block $k$, i.e., $\phi_k=0$, have no effect on total information coding length,
the final formalization of Equation (\ref{SSBM loss function}) can be rewritten by defining $K_{ne} \textless K$ to denote the number of nonempty blocks:

\begin{equation}
\begin{split}
\label{last SSBM loss function}
C(N,g) = -\log p(N|g) +\frac{K_{ne}(c + 1)}{2}\log n \\
+\frac{c}{2}\sum_{\phi_{k} > 0}\log \phi_{k} + \frac{d}{2}(1 + \log \kappa_d)
\end{split}
\end{equation}
\subsubsection{CEM-based Optimization}
In this subsection, the CEM algorithm is adopted to learn model parameters in Equation (\ref{last SSBM loss function}).
Obviously, the right side of Equation (\ref{last SSBM loss function}) is opposite number of the sum of log likelihood, i.e., $\log p(N|g)$, and model prior, i.e., $-\frac{K_{ne}(c + 1)}{2}\log n - \frac{c}{2}\sum_{\phi_{k} > 0}\log \phi_{k} - \frac{d}{2}(1 + \log \kappa_d)$, which is equal to model posterior.
The original optimization task can be transformed from minimizing the cost function to maximizing the posterior.
Specifically, the learning algorithm consists of the following two steps:

\textbf{E-step:} When $N$ and $o^{(t-1)}$ are known, where $o$ represents $(\Phi, \Lambda)$ and $t$ is the times of iterations, the expectation of complete data log likelihood, i.e., the Q function:
\begin{equation}
\begin{split}
\label{EQ23}
Q(o,o^{(t-1)}) = \mathbb{E}_Z[\log p(N,Z|K,o^{(t-1)})]\;\;\;\;\;\;\;\;\;\;\;\;\;\;\;\;\;\;\;\;\;\;\;\;\;\;\;\;\;\;\\
\!\!\!\!=\!\!\sum_{i = 1}^{n}\sum_{k = 1}^{K} \zeta_{ik}\log \phi_{k} \!\!+ \!\!\sum_{i = 1}^{n}\sum_{k = 1}^{K}\sum_{j = 1}^{n}\sum_{h = 1}^{3}\zeta_{ik}\delta(a_{ij},2\!\!-\!\!h)\log \lambda_{kjh}\;\;\;\; %\nonumber
\end{split}
\end{equation}
where $\zeta_{ik}=\mathbb{E}[z_{ik};o^{(t-1)}]$ is the posterior probability that  node $i$ is allocated to block $k$ when $o^{(t-1)}$ is known, and can be calculated as follows:
\begin{equation}
\begin{split}
\label{EQ24}
\zeta_{ik} = \frac{\phi_{k}^{(t-1)}\prod_{j=1}^{n}\prod_{h=1}^{3}\lambda_{kjh}^{\delta(a_{ij},2-h)}}{\sum_{l=1}^{K}\phi_{l}^{(t-1)}\prod_{j=1}^{n}\prod_{h=1}^{3}\lambda_{ljh}^{\delta(a_{ij},2-h)}}
\end{split}
\end{equation}

\textbf{M-step:} Maximizing $Q(o,o^{(t-1)})+\log p(o)$, where $\log p(o) = -\frac{K_{ne}(c+1)}{2}\log n - \frac{c}{2}\sum_{k:\phi_k > 0}\log \phi_{k} - \frac{K_{ne}(c+1)}{2}(1 + \log \kappa_d)$. Since $\sum_{k=1}^{K}\phi_{k}=1$, the Laplace function to be maximized is as follows:
\begin{equation}
\begin{split}
\label{Laplace function}
J =
Q(o,o^{(t-1)})+\log p(o)
+ \eta(\sum_{k=1}^{K}\phi_{k}-1)\\
\end{split}
\end{equation}

Then an explicit solution on $\phi$ can be obtained by calculating the partial derivative of Equation (\ref{Laplace function}).
\begin{equation}
\begin{split}
\label{omega}
\phi_{k}^{(t)} = \frac{\max \{0,\sum_{i=1}^{n}\zeta_{ik} - \frac{c}{2}\}}{\sum_{l =1}^{K}\max \{0,\sum_{i=1}^{n}\zeta_{il}-\frac{c}{2}\}}
\end{split}
\end{equation}

Finally, the probability generating  a positive, negative, or null edge from block $k$ to node $j$, i.e., $\lambda_{kj1}$, $\lambda_{kj2}$, and $\lambda_{kj3}$, can be represented as follows:
\begin{equation}
\begin{split}
\label{theta}
\lambda_{kj1} = \frac{\sum_{i=1}^{n} \zeta_{ik} \delta(a_{ij},1)}{\sum_{i=1}^{n}\zeta_{ik}}\\
\lambda_{kj2} = \frac{\sum_{i=1}^{n} \zeta_{ik} \delta(a_{ij},-1)}{\sum_{i=1}^{n}\zeta_{ik}}\\
\lambda_{kj3} = \frac{\sum_{i=1}^{n} \zeta_{ik} \delta(a_{ij},0)}{\sum_{i=1}^{n}\zeta_{ik}}\\
\end{split}
\end{equation}

In the SSBM model, the ability of model selection can be reflected by Equation (\ref{omega}).
Specifically, $\zeta_{ik}$ in the numerator is posterior probability allocating node $i$ to block $k$,
then $\sum_{i=1}^{n}\zeta_{ik}$ can be regarded as the number of nodes allocated to block $k$.
When the numerator is 0, i.e., $\sum _{ i=1 }^{ n } \zeta _{ ik }<\frac { c }{ 2 } $, block $k$ will be discarded due to be empty, and the  parameters are not estimated in subsequent iterations.
For SSBM, all blocks are empty, and the algorithm is invalid when $c$ is equal to 3$n$.
However, concerning on the right side of Equation (\ref{last SSBM loss function}), it contains the log form of model posterior, which consists of a log form of the Dirichlet prior of $\phi_k$, i.e., $\frac{c}{2}\sum_{\phi_{k} > 0}\log \phi_{k}$, and a log likelihood, i.e., $\log p(N|g)$, when neglecting constant term.
%The Dirichlet-type prior of parameter $\omega_k$ can be represented as:
%
\begin{equation}
\label{EQ30}
{p(\phi_1,\dots,\phi_K)\propto exp\{-\frac{c}{2}\sum_{k=1}^{K} log\phi_k \} = \prod_{k=1}^{K}\phi_k^{-\frac{c}{2}}}
\end{equation}

In Equation (\ref{EQ30}), $-\frac{c}{2}$ is a parameter of the Dirichlet prior, and it has little influence on detecting results of network structure when the data of observable signed network is sufficient.
That is, varying $c$ in Equation (\ref{omega}) will not have a significant impact on the posterior of $\Phi$.
Moreover, the standard SBM needs $K$ parameters to characterize a block, while SSBM requires $3n$ parameters.
The SSBM is essentially an extension of the standard SBM from unsigned networks to signed networks, and
will degenerate to the standard SBM regardless of the signs of edges, and then the number of parameters of each block in SSBM is also $K$.
Therefore, $c/2 = K_{ne}$, i.e., the number of nonempty blocks, can be set in the real applications.

Note that an important heuristic information on setting the upper bound of detectable block space, i.e., $K_{max}$, can be inferred by Equation (\ref{omega}). Seen from the numerator, all the retained blocks in the process of model selection meet $\sum_{i=1}^{n}\zeta_{ik}>K_{ne}$, then  $\sum_{k=1}^{K_{ne}}\sum_{i=1}^{n}\zeta_{ik}>\sum_{k=1}^{K_{ne}}K_{ne}$, and  $K_{ne}<\sqrt{n}$ can be acquired, which can be seemed as the resolution limit of SSBM. Therefore, when there is no prior knowledge on signed networks, the heuristic information can provide a good choice for initializing the maximum detectable block number, i.e., $K_{max}=\sqrt{n}$.

\subsection{Time Complexity Analysis}

\begin{figure*}[htbp]
  \centering
    \includegraphics[width=0.9 \textwidth]{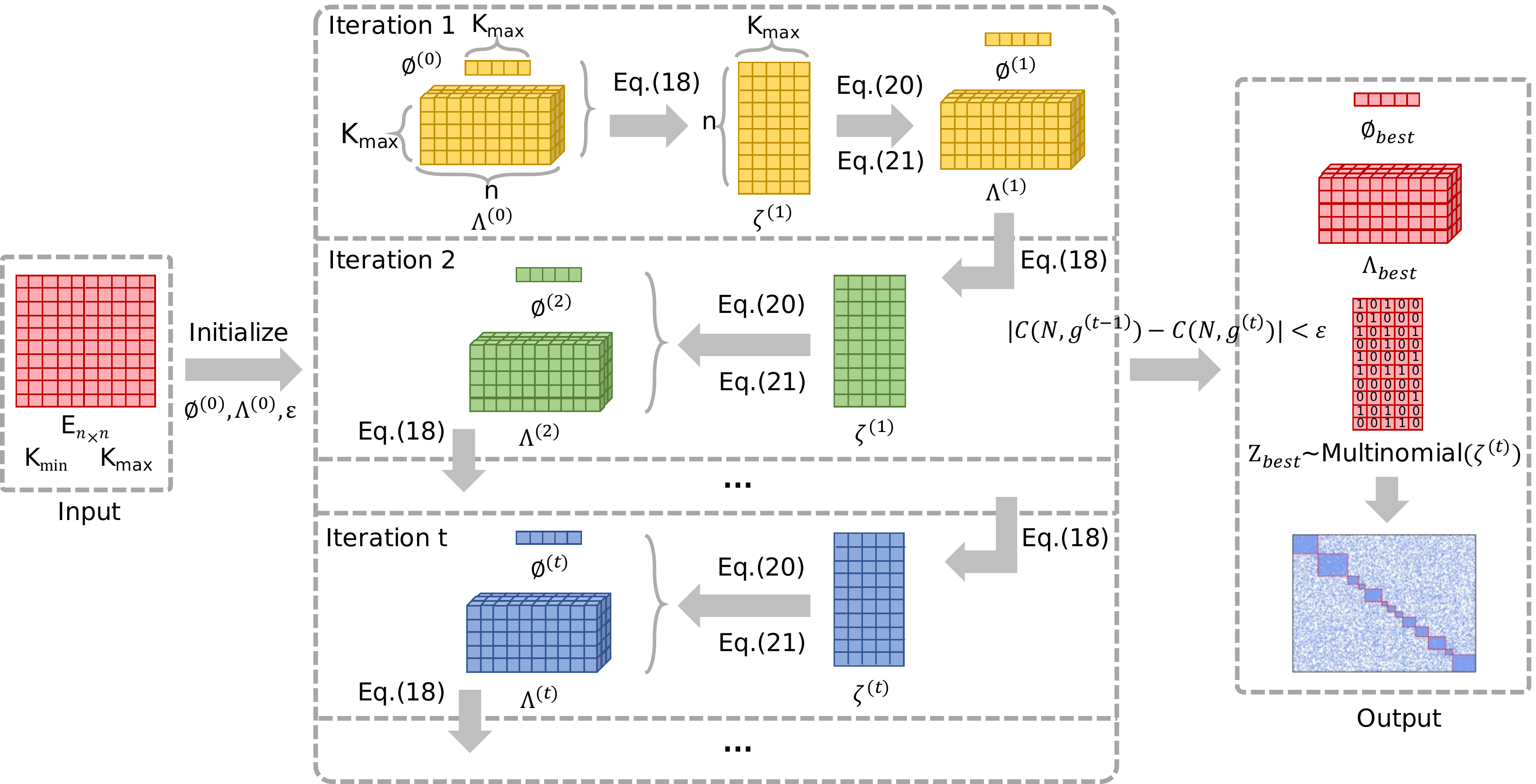}
  \captionsetup{justification=centering}
  \caption{\normalsize{The visual flow chart of SSBM.}}\label{flowchart}
  \label{synthetic networks}
\end{figure*}

%The details of SSBM learning algorithm are shown in Table 1.
%In order to show more intuitively that our proposed algorithm can synchronously perform parameter estimation and model selection, necessary annotations are given in Table 1.
The pseudo code of SSBM learning algorithm is presented in Algorithm \ref{alg:Framwork}, and the flow of execution is visualized in Fig. \ref{flowchart}. Obviously, the most time-consuming parts of SSBM learning algorithm are repeat loop and foreach loop.
The foreach loop is responsible for evaluating block, selecting block, estimating parameters, and discarding block.
The repeat loop repeats above calculations until cost function converges, and then the optimal model and latent variable $Z$ are obtained (Lines 24 and 25).
Specifically, in foreach loop, $\zeta^{(t)}_{ik}$ is calculated first, and then $\phi^{(t)}_{k}$ (Lines 7 to 9).
Since $\zeta^{(t)}_{ik}$ is posterior probability after the $t^{th}$ iteration,
the block quality can be evaluated based on $\sum _{ i=1 }^{ n }{ { \zeta  }_{ ik }^{ (t) } } $ and $\frac { c }{ 2 } $.
For example, if $\sum _{ i=1 }^{ n }{ { \zeta  }_{ ik }^{ (t) } } <\frac { c }{ 2 } $, i.e., block $k$ cannot be supported by data, $\phi^{(t)}_{k} = 0$ and block $k$ will be discarded (Line 15); otherwise, block $k$ will be selected (Line 10). Then the parameters of selected block will be estimated (Lines 11 and 12).

%The parallel learning algorithm of the SSBM is detailed in Algorithm \ref{alg:Framwork}.
The time complexity of calculating $\zeta_{\cdot k}$ and $\phi_k$ in Lines 7 and 8 are $O(nK_{max})$ and $\lambda_{\cdot k}$ and $u_{\cdot k}$ in Lines 11 and 12 are $O(n^2)$, respectively.
Thus, the time complexity of executing a complete foreach loop is $O(n^2K_{max}+nK^2_{max})$.
The time complexity of calculating $C(N,g^{(t)})$ is $O(n^2K_{max}+K_{max})$.
The above calculations cost $O(2n^2K_{max}+nK^2_{max}+K_{max})$ in total.
Assuming that the algorithm converges after $T$ iterations, the repeat loop is $O(Tn^2K_{max})$. Therefore,
the time complexity of SSBM is $O(Tn^2K_{max}(K_{max}-K_{min}))$ due to the while loop executing $K_{max}-K_{min}$ times.

\begin{algorithm}[t]
\caption{SSBM Learning Algorithm}
\label{alg:Framwork}
\LinesNumbered
\KwIn{$N,K_{min},K_{max}$}
\KwOut{$g_{best},Z_{best}$}
Initialize $\Phi^{(0)};\Lambda^{(0)};t\leftarrow0;K_{ne} \leftarrow K_{max}; \varepsilon; $ \\
    $u_{ik}^{(0)}\leftarrow\prod_{j=1}^n\prod_{h=1}^3(\lambda_{kjh}^{(0)})^{\delta(a_{ij},2-h)},$ for  $i=1,..,n$ and $k=1,..,K_{max};$
    $C_{min}\leftarrow+\infty;g^{(0)}=(K_{max},\Phi^{(0)},\Lambda^{(0)})$

\While{$K_{ne} \geqslant K_{min}$}{
\bf{repeat}\\
$t\leftarrow t+1;$ \\
\ForEach {$k=1$ to $K_{max}$}{
     $\zeta_{ik}^{(t)}\leftarrow \frac{\phi_k^{(t-1)}u_{ik}^{(t-1)}}{\sum_{l=1}^{K_{max}}\phi_l^{(t-1)}u_{il}^{(t-1)}},$ for $i=1,..,n;$ \\
     $\phi_k^{(t)}\leftarrow \frac{max\{0,\sum_{i=1}^n\zeta_{ik}^{(t)}- \frac{c}{2}\}}{\sum_{l=1}^{K_{max}}max\{0,\sum_{i=1}^n\zeta_{il}^{(t)}-\frac{c}{2}\}};$ \\
    $\phi_q^{(t)}\leftarrow \phi_q^{(t)}(\sum_{l=1}^{K_{max}}\phi_l^{(t)})^{-1};$\\
    \If{$\phi_k^{(t)}>0$ }{
    $\lambda_{kjh}^{(t)}\leftarrow\frac{\sum_{i=1}^n\zeta_{ik}\delta(a_{ij},2-h)}{\sum_{i=1}^n\zeta_{ik}},$ for $j=1,..,n,$  $h=1,2,3;$ \\
    $u_{ik}^{(t)}\leftarrow\prod_{j=1}^n\prod_{h=1}^3(\lambda_{kjh}^{(t)})^{\delta(a_{ij},2-h)}, $ for $i=1,..,n;$ \\}
    \If{$\phi_k^{(t)}\leq0$ }{$K_{ne}\leftarrow K_{ne}-1;$}
    }
    $g^{(t)}=(K_{ne},\Phi^{(t)},\Lambda^{(t)});$ \\
    $C(N,g^{(t)})\leftarrow\frac{K_{ne}(c+1)}{2}\log n+\frac{c}{2}\sum_{\phi_k>0}\log\phi_k +\frac{K_{ne}(c+1)}{2}(1+\log\kappa_d)-\sum_{i=1}^n\log\sum_{k=1}^{K_{max}}\phi_k^{(t)}u_{ik}^{(t)};$ \\
    \bf{until}\\
    $(C(N,g^{(t-1)})-C(N,g^{(t)}))<\varepsilon$ \\
    \If {$C(N,g^{(t)})<C_{min}$}{
    $C_{min}\leftarrow C(N,g^{(t)});$ \\
    $g_{best}\leftarrow g^{(t)};$ \\
    $Z_{best}\sim multinomial(\zeta^{(t)});$ \\
    }
}
%\textbf{Return} $\mathbf{\Phi}$,$\mathbf{A}$, $\mathbf{B}$, $\mathbf{C}$.
\end{algorithm}

\section{Validation}
In this section, the generalization, robustness and scalability of SSBM will be validated by comparisons with state-of-the-art methods on synthetic and real-world networks.
The programs of all algorithms are developed using MATLAB 2010b and run on a computer with a 4-core CPU with a 3.20, 8GB RAM, and 64-bit Windows 10 operating system. 
For all the compared algorithms, $K_{min} = 1$ and $K_{max}=10$ are set uniformly, and for SSBM, $K_{min} = 1$ and $K_{max}=\sqrt{n}$ are set in term of the aforementioned heuristic information, and $\Phi$ is initialized by ${(1/K_{max},\dots,1/K_{max}})$, $\Lambda$ is initialized randomly, $\epsilon=10^{-4}$.

\subsection{Metrics and Baselines}
The accuracy of signed network structure discovery is evaluated by normalized mutual information (NMI) \cite{Kuncheva}.
Assume that $A$ and $B$ are the real and discovered network structure partitions, respectively. The NMI can be calculated as:
\begin{equation}
\label{NMI}
NMI(A,B)=\frac { -2\sum _{ i=1 }^{ { C }_{ A } }{ \sum _{ j=1 }^{ { C }_{ B } }{ { m }_{ ij }\log { (\frac { { m }_{ ij }M }{ { m }_{ i\cdot  }{ m }_{ \cdot j } } ) }  }  }  }{ \sum _{ i=1 }^{ C_{ A } }{ { m }_{ i\cdot  }\log { (\frac { { m }_{ i\cdot  } }{ M } ) } +\sum _{ j=1 }^{ C_{ B } }{ { m }_{ \cdot j }\log { (\frac { { m }_{ \cdot j } }{ M } ) }  }  }  } \nonumber
\end{equation}
where $M$ is confusion matrix, and $m_{ij}$ denotes the number of nodes belonging to block $i$ of $A$ and block $j$ of $B$ at the same time.
$C_A$ and $C_B$ are the number of blocks in $A$ and 
$B$, respectively.
$m_{\cdot i}$ and $m_{\cdot j}$ represent the sum of the elements in Row $i$ and Column $j$ in $M$, respectively.
If the discovered partition $B$ is identical to the real partition $A$,   $NMI(A,B)=1$; otherwise, $NMI(A,B)=0$.

Five classical signed network structure
methods, namely, VBS \cite{Zhao_and_Liu}, SSL \cite{Yang_and_Liu}, SISN \cite{Zhao_and_Yang}, FEC \cite{Yang_and_Cheung}, and DM \cite{Doreian_and_Mrvar}, are selected as compared methods.

\begin{itemize}
\item \emph{VBS} \cite{Zhao_and_Liu} is a learning algorithm with model selection, which is proposed by extending the standard SBM to signed SBM in the framework of variational Bayesian.

\item \emph{SSL} \cite{Yang_and_Liu} is a variational Bayesian EM algorithm based on approximate evidence, and a signed stochastic block model is defined by explicitly modeling the density and noise distribution of the edges.
%for parameter estimation and model selection

\item \emph{SISN} \cite{Zhao_and_Yang} is a signed network community discovery method based on statistical reasoning, and 
a model selection strategy based on the minimum description length (MDL) is presented.

\item \emph{FEC} \cite{Yang_and_Cheung} is a method based on random walk model to discover signed network communities by alternately executing FC (finding a municipality) and EC (extracting a community).
%According to the predefined graph cut criteria, EC extracts sink community from the current signed network.

\item \emph{DM} \cite{Doreian_and_Mrvar} is a local search method based on equilibrium theory proposed by Doreian and Murvar, which detects networks structures by minimizing the noise of signed networks.
\end{itemize}

\subsection{Validation on Synthetic Signed Networks}
In this section, the effectiveness of SSBM will be validated using synthetic networks compared with five state-of-the-art methods.

\subsubsection{Synthetic Datasets Generation}

For the sake of fairness, synthetic signed networks are mostly generated using the model proposed in \cite{Yang_and_Cheung}:
\begin{equation}
\label{generative model}
Model_{sign}=SG(c,m,k,p_{in},p-,p+) %\nonumber
\end{equation}
where $c$, $m$, and $k$ are the number of blocks, the number of nodes in a block, and average node degree, respectively.
$p_{in}$ is a parameter for controlling cohesiveness.
It represents the probability of generating edges between nodes within blocks.
The closer the value is to 1, the more distinct the network structure.
$p-$ and $p+$ are parameters for controlling the noise.
They represent the probabilities of generating negative edges within blocks and positive edges between blocks, respectively. The larger the two values, the more complex the network structure.
\begin{figure*}[htbp]
  \centering
    \includegraphics[width=0.90 \textwidth]{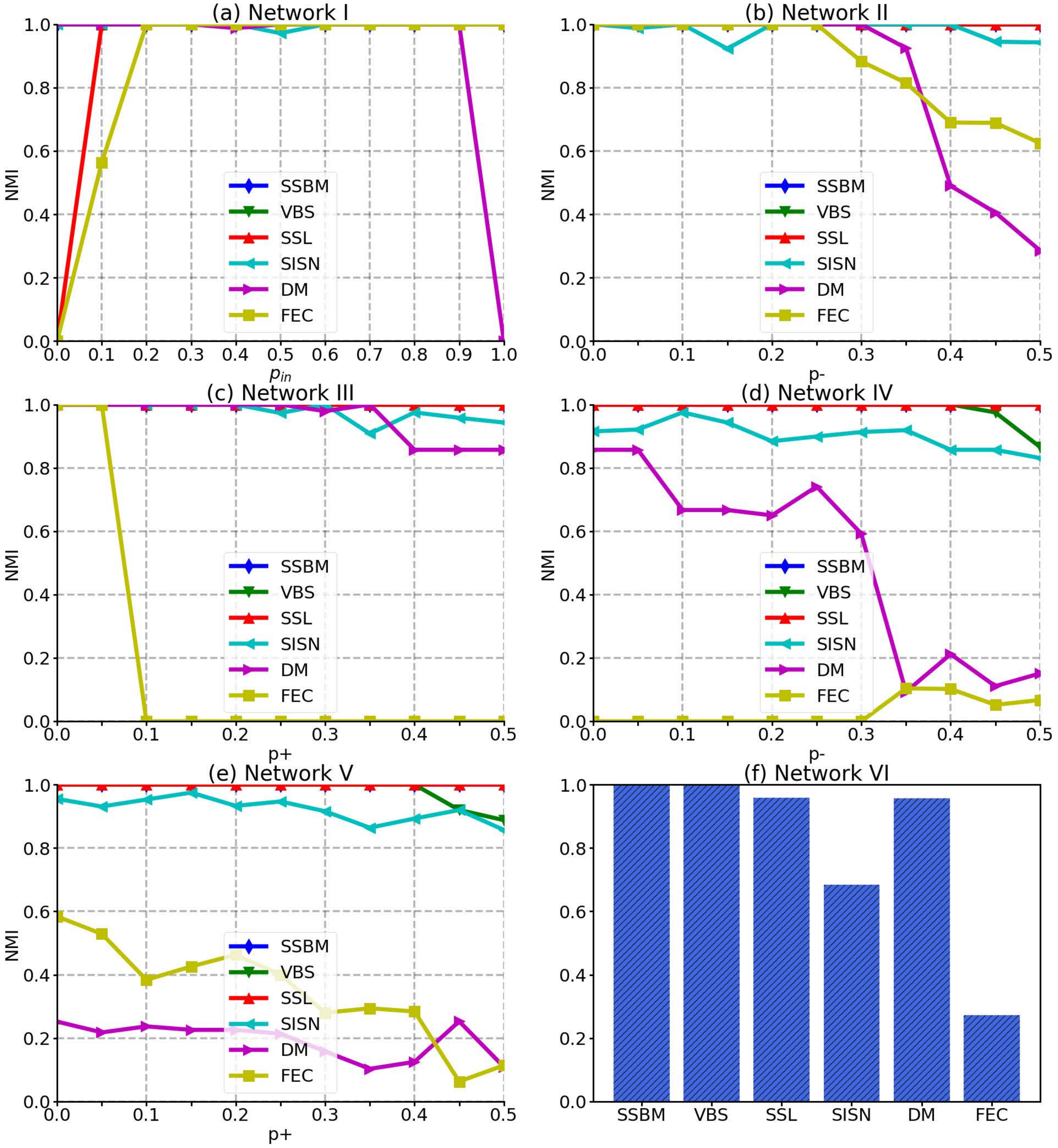}
  \captionsetup{justification=centering}
  \caption{\normalsize{Experimental results of six algorithms on six kinds of synthetic networks.}}
  \label{synthetic networks}
\end{figure*}

Five kinds of synthetic networks are generated using the above generative model.
%Five kinds of synthetic signed networks are generated utilizing the model introduced in equation (28) to verify the effectiveness of the SSBM.

\begin{itemize}
\item \emph{Network I}: It is generated by parameter configuration (4,32,32,$p_{in}$,0,0),
where $p_{in}$ increases from 0.0 to 1.0, and the step size is 0.1.
It is a balanced signed network where  positive edges only exist within blocks while negative edges only exist between blocks.

\item \emph{Network II}: It is generated by parameter configuration (4,32,32,0.6,p-,0),
where $p-$ increases from 0.0 to 0.5, and the step size is 0.05.
The negative edges within blocks can be seen as network noises.
As $p-$ increases, there are more negative edges within blocks.

\item \emph{Network III}: It is generated by parameter configuration (4,32,32,0.6,0,p+),
where $p+$ increases from 0.0 to 0.5, and the step size is 0.05.
The positive edges between blocks can be seen as noises.
As $p+$ increases, there are more positive edges between blocks.

\item \emph{Network IV}: It is generated by parameter configuration (4, 32, 32, 0.6, p-, 0.5),
where $p-$ also increases from 0.0 to 0.5, and the step size is also 0.05.
There are two kinds of noises in the generated network, i.e., positive edges between blocks and negative edges within blocks, and it is more complex than network II.

\item \emph{Network V}: Similar to Network IV, it is generated by parameter configuration (4, 32, 32, 0.6, 0.5, p+), where negative edges within blocks are generated with the probability of 0.5, and positive edges between blocks are generated according to $p+$.
\end{itemize}

Networks II-V are unbalanced signed networks that are mainly used to validate the robustness of the proposed method.
To further verify the ability of SSBM to discover multiple structures, Network VI containing community and bipartite structures is generated by the following method.

\begin{itemize}
\item \emph{Network VI}: It is generated by dividing all nodes into two communities and two bipartites, and each block contains 32 nodes.
The number of positive, negative, or null edges generated within and between blocks is subject to multinomial distribution with parameters of $\pi_{kk}$ and $\pi_{kq}$, respectively.
Specifically, $\pi_{kk}$ and $\pi_{kq}$ are set as follows:

$\pi_{11}=\{0.6,0.1,0.3\}, \pi_{12}=\{0.1,0.2,0.7\}, \pi_{13}=\\ \{0.1,0.2,0.7\}, \pi_{14}=\{0.1,0.2,0.7\}$; \\
$\pi_{22}=\{0.2,0.1,0.7\}, \pi_{23}=\{0.01,0.4,0.59\}, \pi_{34}=\\ \{0.01,0.4,0.59\}$; \\
$\pi_{33}=\{0.01,0.01,0.98\}, \pi_{34}=\{0.01,0.4,0.59\}$; \\
$\pi_{44}=\{0.01,0.01,0.98\}$.
\end{itemize}

\subsubsection{Results and Analysis}

Fig. \ref{synthetic networks} presents the experimental results on synthetic networks.
For balanced signed networks (Fig. \ref{synthetic networks} (a)), SSBM and VBS show excellent performance in community structure discovery. 
With an increase of $p_{in}$ from 0 to 1, both of them can find communities accurately, indicating that they are insensitive to cohesiveness within blocks.
The accuracy of SISN decreases slightly only when $p_{in}=0.5$, and SSL and DM are worse than SISN.
SSL is incapable of discovering community structure when $p_{in}$ = 0 since the cohesiveness in the community is very weak, while with an increase of $p_{in}$, the community discovery ability of SSL is gradually enhanced. 
In contrast to SSL, when cohesiveness in the community is very strong, e.g., $p_{in}=1$, DM cannot discover community structure. 
When $p_{in}<1$, DM has a good ability of community discovery.
%FEC is the worst one.

Experimental results on five kinds of unbalanced signed networks are presented in Fig. \ref{synthetic networks} (b)-(f), respectively. 
In Fig. \ref{synthetic networks} (b), SSBM, VBS, and SSL perform the best. They are not affected by noises within communities, and can accurately find community structure.
SISN can detect the community structure  in most cases with noise ratio $p-$ varying, and it achieves the best performance (NMI = 0.923) when $p-=0.15$.
For FEC and DM, the accuracy declines rapidly when $p-$ increases gradually,
indicating that they are more sensitive to noises within communities.
In Fig. \ref{synthetic networks} (c), SSBM, VBS, and SSL are also the best ones, and they can accurately discover communities in all cases.
As $p+$ increases, the performance of SISN and DM decreases slightly, but the NMIs are no less than 0.9 and 0.85, respectively. 
FEC is extremely sensitive to noises between communities, and it can accurately find the communities only when $p+<0.1$; otherwise, it becomes ineffective.

As shown in Fig. \ref{synthetic networks} (d) and (e), SSBM and SSL still exhibit competitive performance. VBS can accurately discover  community structure when $p+\le0.4$, and it decreases slightly with an increase of $p+$. SISN is slightly worse than VBS, but the NMI is always larger than 0.83.
FEC and DM are the worst. Specifically, for Network IV, DM performs better than FEC. FEC is invalid when $p-\le0.3$.
For Network V, FEC is better than DM. Experimental results on Network VI, which contains both community and bipartite structures, are shown in Fig. \ref{synthetic networks} (f). Specifically, the NMIs of SSBM, VBS, SSL, SISN, DM, and FEC are 1, 1, 0.958, 0.684, 0.957, and 0.273, respectively.

In summary, SSBM consistently exhibits the excellent performance for all the synthetic signed networks.
Results on unbalanced signed networks further demonstrate that SSBM has good robustness to different network noise types and intensities, and shows strong generalization as well, i.e., it can accurately find multiple structures existed in the networks. Because real-world networks usually contain various noises and  structures,
SSBM has more advantages on discovering multiple structures in real-world networks  than compared methods.
\begin{table*}[htbp]
\centering
\caption{\normalsize{Accuracy on three real-world signed networks}} 
\begin{tabular}{cccccccc}
\toprule
Networks   &$K_{true}$   &SSBM    &VBS     &SSL     &SISN     &DM    &FEC\\
\hline
SPPN   &2  &\textbf{1}/\textbf{2}   &\textbf{1}/\textbf{2}   &\textbf{1}/\textbf{2}    &\textbf{1}/\textbf{2}   &\textbf{1}/\textbf{2}  &0.619/\textbf{2}\\
GGSN  &3    &\textbf{1}/\textbf{3}    &\textbf{1}/\textbf{3}     &\textbf{1}/\textbf{3}      &0.528/\textbf{3}    &0.938/4   &0.911/4\\
MN   &3  &\textbf{1}/\textbf{3}     &0/1  &0/1   &\textbf{1}/\textbf{3}      &0.86/\textbf{3}  &0.464/\textbf{3}\\
\hline 
Avg(rank)   &    &\textbf{1}(\textbf{1}) &0.667(4) &0.667(4) &0.843(3)    &0.933(2)  &0.665(5) \\
\bottomrule
\end{tabular} \label{NMIonrealworldnetworks}
\end{table*}

\subsection{Validation on Real-World Signed Networks}
In this section, the ability of SSBM is further validated by dealing with real-world networks.

\subsubsection{Description of Real-World Datasets}

In this experiment, we select three real-world datasets with ground truth, i.e., the Slovene Parliamentary Party Network (SPPN) \cite{Kropivnik}, the Gahuku-Gama Subtribes Network (GGSN) \cite{Read}, the Monastery Network (MN) \cite{Sampson}, and a real-world dataset without ground truth, i.e., Country Network \cite{Doreian}, to validate the effectiveness of SSBM.
\begin{itemize}
\item
SPPN \cite{Kropivnik} is a signed network representing the relations among political parties in the Slovenian Parliament in 1994.
The network contains 10 nodes, 2 communities , and 18 positive edges and 27 negative edges indicating that these political parties have similar or opposite political relations.
\item
GGSN \cite{Read} describes political relationships between the Gahuku-Gama subtribes in 1954.
The network contains 16 nodes, 3 communities, and 29 positive edges and 29 negative edges representing alliances or hostile political relations between political parties.
\item
MN \cite{Sampson} describes the emotion relations, i.e., like or dislike, among monks in the New England monastery.
There are 18 nodes, 3 communities, and 51 positive edges and 58 negative edges.
\end{itemize}

\subsubsection{Results and Analysis}

The results on three real-world networks with ground truth are shown in Table \ref{NMIonrealworldnetworks}.
$K_{true}$ denotes the number of true communities, the value before ``/'' represents the NMI, and the value after ``/'' represents the number of communities discovered by algorithms.
In the last line, the average NMI of each algorithm is calculated, and the value in ``()'' represents the ranking of each algorithm based on average NMI.
SSBM outperforms all the compared methods because it can accurately discover network structures for three real-world networks. 
This further demonstrates that SSBM, using the block-to-node connection probability matrix $\Lambda$ to reparameterize $\Pi$ in the standard SBM, can capture more fine-grained network structure information and effectively improve the accuracy of multiple structure discovery for signed networks.

\begin{figure}[htbp]
  \centering
  \includegraphics[width=0.41 \textwidth]{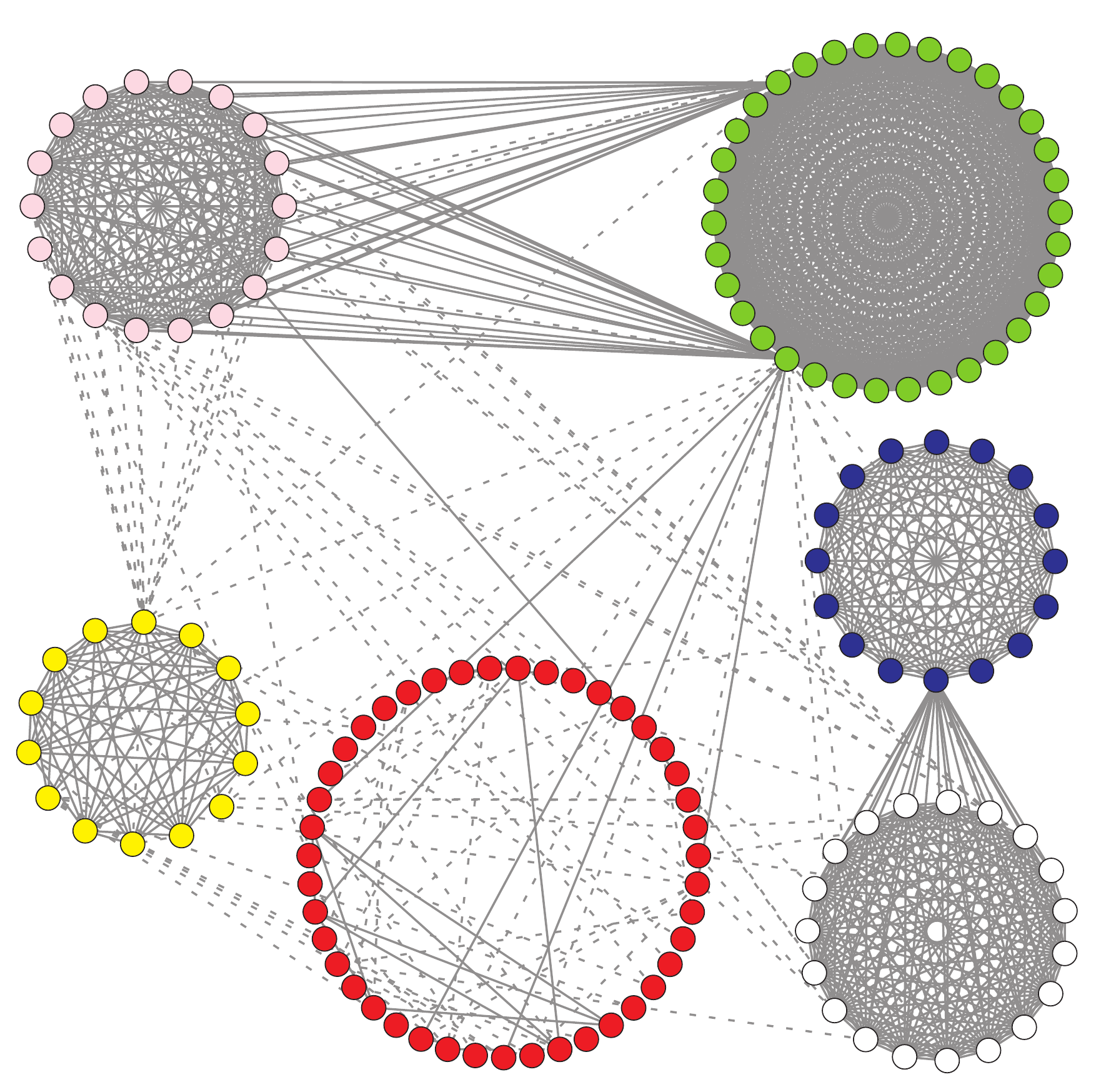}
  \caption{\normalsize{The partition visualization of SSBM for Country Network.}}
  \label{visual result}
\end{figure}

SSBM is further validated using a real-world dataset without ground truth, namely, the Country Network. It is derived from the Correlates of War dataset of countries from 1996 to 1999.
After removing isolated nodes and sparsely connected components, the Country Network ultimately contains 144 nodes, 1099 positive and 144 negative edges, representing military alliances and military disputes, respectively.
The partition result given by SSBM on the Country Network is shown in Fig. \ref{visual result}.
Each color represents a cluster, and the solid dots represent nodes of the Country Network, and the positive and negative edges are denoted by solid and dotted gray lines, respectively.
SSBM divides the Country Network into six clusters, where the number of nodes is 14 (yellow), 16 (blue), 18 (pink), 19 (white), 34 (green), and 43 (red), respectively.
Obviously, this partition is quite reasonable: 
positive edges are mainly distributed within blocks, while negative edges mainly exist between blocks, and there are basically no controversial nodes.

The above validations show that SSBM has excellent accuracy and good generalization ability. 
It can detect the block structures in the signed networks reasonably and efficiently without any prior knowledge, and is more applicable for handling real-world exploratory signed networks.

\subsection{Validation of Scalability}

\begin{figure*}[htbp]
  \centering
    \includegraphics[width=0.90 \textwidth]{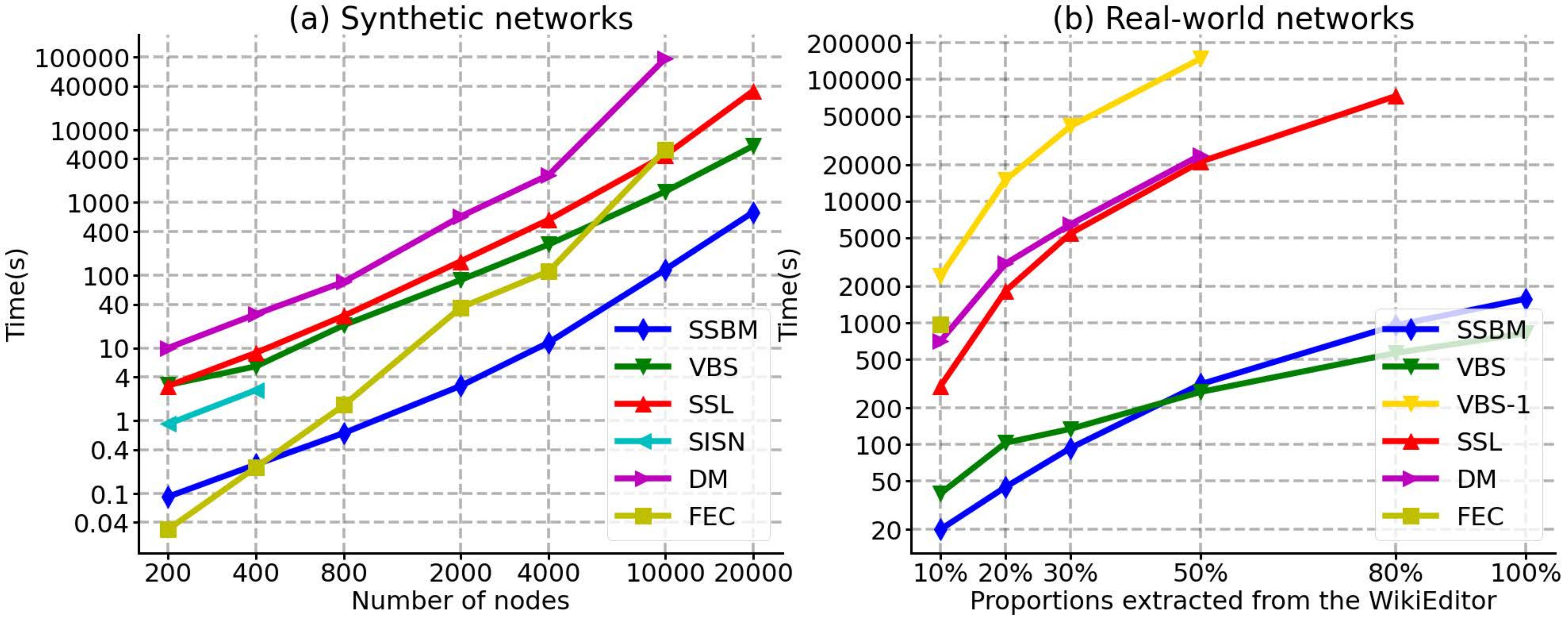}
  \captionsetup{justification=centering}
  \caption{\normalsize{Running time on synthetic and real-world large-scale signed networks.}}
  \label{running time}
\end{figure*}

For validating scalability , a series of unbalanced synthetic signed networks with different scales are generated using the aforementioned model (Equation (\ref{generative model})).
All of the unbalanced synthetic networks contain four clusters, i.e., $c$ in Equation (\ref{generative model}) is 4. The $m$ and $k$ are set as $50$, $100$, $200$, $500$, $1000$, $2500$, and $5000$ in sequence. That is,
the numbers of nodes $n$ are $200$, $400$, $800$, $2000$, $4000$, $10000$, and $20000$, respectively. For all networks, $p-=0.5$ and $p+=0.5$. When $n\le10000$, $p_{in}=0.8$; otherwise, $p_{in}=0.4$.
Experimental results on accuracy and running time are presented in Table \ref{tab:scaleNMI} and Fig. \ref{running time} (a), respectively.
As seen from Table \ref{tab:scaleNMI}, SSBM and SSL are the best ones in all cases because the NMIs of  two methods are always 1. VBS is slightly worse than SSBM and SSL, and it can discover network structures accurately in most cases. SISN is only capable of dealing with small networks. DM and FEC are worst, and the results indicate that the two methods are unadaptable to the signed networks containing complex noises.
\begin{table*}[htbp]
\centering
\caption{\normalsize{Accuracy on large-scale synthetic signed networks}} 
\begin{tabular}{cccccccc}
\toprule
Nodes   &200 &400  &800  &2000 &4000 &10000 &20000 \\
\hline
SSBM       &\textbf{1}      &\textbf{1}   &\textbf{1}  &\textbf{1} &\textbf{1} &\textbf{1}  &\textbf{1}\\
VBS       &\textbf{1}     &\textbf{1}        &\textbf{1}  &\textbf{1} &0.857 &0.857  &\textbf{1}\\
SSL       &\textbf{1}     &\textbf{1}        &\textbf{1}  &\textbf{1} &\textbf{1}  &\textbf{1}  &\textbf{1}\\
SISN      &\textbf{1}      &\textbf{1}        &--  &--  &-- &-- &--\\
DM      &0.04     &0.007        &0.022  &0.004 &0.003 &0.0006 &--\\
FEC      &0.01    &0.08        &0.026 &0.012 &0 &0.003  &-- \\
\bottomrule
\end{tabular} \label{tab:scaleNMI}
\end{table*}

In Fig. \ref{running time} (a), although SSBM and SSL have the same excellent accuracy, SSL needs a significant amount of time. 
For example, when $n$ is $200$, SSBM and SSL take $0.09$ seconds and $3.0$ seconds, respectively. 
As network scale increases, the advantage of SSBM on dealing with large-scale networks is highlighted. 
When $n$ is $10000$, the running times of SSBM and SSL are $120$ seconds and $4467.8$ seconds, respectively. 
That is, SSBM is $37$ times more efficient than SSL. Furthermore, when $n$ is $20000$, the performance gap between the two methods is even greater. 
SSBM only takes $739.2$ seconds to discover network structures accurately, while SSL requires $34463.3$ seconds. 

Moreover, the experiments on real-world signed networks are designed to further the scalability of SSBM. The WikiEditor is one of the few real-world signed networks with a slightly larger scale \cite{KumarKDD2015}. It contains 21535 nodes, 269251 positive edges and 79004 negative edges. The nodes represent users participating in editing  Wikipedia pages from Jan 2013 to July 2014, who were classified into benign users and vandals. Each edit of any user can belong to either revert or no-revert category. The edges between nodes are built on co-edit relations, that is, if most of the co-edits for two users belong to the same category, there is a positive edge between the two users; otherwise, there is a negative edge. To verify the scalability, a set of datasets are generated by  extracting from the WikiEditor in proportions as $10\%$, $20\%$, $30\%$, $50\%$, $80\%$ and $100\%$, respectively. 
The running times of all the algorithms on the datasets are shown in Fig. \ref{running time} (b).

SISN has the worst performance, and even cannot deal with the dataset composed of $10\%$ of the WikiEditor. Therefore, it is not exihibited in Fig. \ref{running time} (b). The second worst is FEC, and it will terminate due to lack of memory when the size of dataset is more than $20\%$ of the WikiEditor. The reason is that the edges of the WikiEditor are relatively dense, and FEC requires more computing resources in the execution process. DM performs better than FEC, and can handle datasets with the size less than half of the WikiEditor, although it takes a long time. SSL performs better than DM. SSL can detect the larger datasets with the size no more than $80\%$ of the wikiEditor, and presents more faster running speed. SSBM and VBS are the best algorithms. Both of them can handle all the datasets in a short time, and when the size of dataset is equal or greater than $50\%$ of the WikiEditor, VBS takes less time than SSBM. For example, when the dataset is the WikiEditor itself, SSBM takes $1572.06$ seconds, while VBS takes only $819.16$ seconds. The result seems to indicate that VBS is more efficient than SSBM in dealing with large-scale real-world signed networks, but it is really not so. This is because that SSBM and VBS are not fair in the settings of model search space. In term of the heuristic information discussed previously, the model search space of SSBM for all experiments is $[1,\sqrt{n}]$, while VBS is $[1,10]$. Therefore, the larger $n$ is , the larger model space SSBM need to search, and the more time it will take. To be fair, a new experiment on setting the model search space of VBS as same as SSBM, i.e., the model search space of VBS is also $[1,\sqrt{n}]$, is presented. The corresponding result is labeled by VBS-1 in Fig. \ref{running time} (b). Obviously, the running time of VBS is much more than SSBM under the same condition. For example, when the size of dataset is $50\%$ of the WikiEditor, SSBM takes $312.09$ seconds while VBS takes $148449.14$ seconds. As the size of dataset  increasing, VBS will be invalid.

The experimental results in term of scalability show that the learning mechanism,  synchronously carrying out parameter estimation and model selection, can significantly improve SSBM learning efficiency
This make SSBM have more advantages and potentials in dealing with large-scale exploratory signed networks, especially unbalanced signed networks containing multiple structures and various noises.

\subsection{Further Discussion}
In this section, the advantages of SSBM in terms of generalization, robustness, and scalability are summarized.
SSBM is essentially an extension of the standard SBM.
First, we use block-to-node connection probability matrix $\Lambda$ to reparameterize block-to-block connection probability matrix $\Pi$. This enables the extended model to capture structure information of networks from a more fine-grained perspective, and  makes it more expressive than the standard SBM.
Second, we let $\Lambda$ be subject to a multinomial distribution, and it can explicitly depict the probabilities of generating positive, negative, and null edges and further exploit the applications of the standard SBM from unsigned networks to signed networks.
These two extensions enable SSBM to accurately discover multiple structures in various signed networks.

The SSBM has good robustness to noises because the parameter $\Lambda$ can explicitly model noise densities in the signed networks.
Taking $\Lambda_{kj}=(\lambda_{kj1}, \lambda_{kj2}, \lambda_{kj3})$ as an example, $\lambda_{kj1}$ and $\lambda_{kj2}$ represent the probabilities that generating positive and negative edges between any node in block $k$ and node $j$, respectively.
If node $j$ belongs to block $k$, then $\lambda_{kj2}$ is the probability of generating a negative edge between any node in block $k$ and node $j$, i.e., intrablock noise, and the expectation $\sum _{ j\in k }^{  }{ \sum _{ i\in k }^{  }{ { \lambda  }_{ kj2 } }  } $ denotes the density of intrablock noises.
Conversely, if node $j$ does not belong to block $k$, then $\lambda_{kj1}$ denotes the probability that a node in block $k$ generates a positive edge to node $j$, and its expectation $\sum _{ j\notin k }^{  }{ \sum _{ i\in k }^{  }{ { \lambda  }_{ kj1 } }  } $ denotes the density of interblock noises.
Therefore, SSBM can effectively overcome the influence of network noises with different types, e.g., negative edges in communities or positive edges between communities, and different densities, e.g., various configurations of $p +$ and $p-$.

According to the above analysis, the potential applications of SSBM can be briefly summarized in the following four perspectives: 
1) discovering and extracting multiple structures in heterogeneous signed networks; 
2) discovering and extracting single or multiple structures in signed networks containing complex noises; 
3) discovering and extracting single or multiple structures in large-scale signed networks; 
4) discovering and extracting single or multiple structures in exploratory signed networks. 

It has to be said that $K_{max}$ is set as $\sqrt{n}$ for SSBM according to the heuristic information, i.e., the block search space is [1, $\sqrt{n}$]. Therefore, when the number of blocks contained in a signed network is greater than $\sqrt{n}$, SSBM cannot correctly detect the network structures. $K_{max}=\sqrt{n}$ is the resolution limit of SSBM. For this case, a strategy can be employed to flexibly set the block search space according to the space complexity $O(\sqrt{n})$ , i.e., set [$K_{min}$, $K_{max}$] as $[(m-1)\sqrt{n}+1, m\sqrt{n}]   (m=1,2,...,\sqrt{n})$. This strategy can ensure that SSBM can accurately discover the network structures,  effectively reduce the search space, and improve the algorithm's efficiency.

\section{Related Work}

This section introduces previous studies closely related to our method from discriminant and principled perspectives.

\subsection{Discriminant Methods}
Discriminant methods usually divide nodes to different clusters based on predefined optimization objectives \cite{Zhu18} or heuristic information \cite{Esmailian15, Chen16}.
In 1996, Doreian and Murvar proposed a frustration-based signed network community discovery method named as DM \cite{Doreian_and_Mrvar}, which detect communities by minimizing network noises.
There are two kinds of noises, i.e., negative edges in communities and positive edges between communities. 
Besides community structure, DM can also detect bipartite structure and coexistence of community and bipartite structures.
In view of this, DM can be seen as the early multiple structure discovery approach. 
Bnasal et al. proposed a community detection method for signed networks by maximizing the sum of positive edges within and  negative edges between communities \cite{Bansal_and_Blum}.
Traag and Bruggeman proposed a modularity-based community partition method by maximizing the modularity of signed networks \cite{Traag_and_Bruggeman}.
In addition, evolutionary computation and non-negative matrix factorization methods are used to detect the community structure.
For instance,
in 2016, Li et al. proposed an optimization method based on a multiobjective particle swarm to discover communities \cite{Li_and_He}.
In 2018, Zhu et al. used evolutionary algorithm to detect community structure of unbalanced signed networks \cite{Zhu_and_Ma}. Recently, Li et al presented a new form of non-negative matrix factorization and a probabilistic surrogate learning function for community detection, but this method can only be applicable to unsigned networks \cite{lihuijia2020}.

Most of discriminant methods mentioned above can only discover community structure of signed networks.
Moreover, the scalability of these methods is very limited.
In view of this, Yang et al. proposed a fast community discovery method FEC by employing a Markov stochastic process \cite{Yang_and_Cheung}.
Anchuri et al. proposed a spectral method to find communities by optimizing modularity or other objective functions \cite{Anchuri_and_Magdon}.
These methods can effectively handle signed networks with thousands of nodes, but they still cannot detect multiple structures and are very sensitive to network noises.
For all of discriminant methods, there is a common problem that their performances overly rely on   predefined optimization objectives or heuristic information.

\subsection{Principled Methods}
For principled methods, network structures can be  discovered from a network generation perspective by fitting probabilistic model to observed  networks \cite{Mucha10, Doreian08}.
For example, Zhao et al. proposed a probabilistic model named as SISN for detecting community structure of signed networks \cite{Zhao_and_Yang}.
Compared with discriminant methods, principled methods can find the intrinsic structures more accurately and have good interpretability, but they also cannot discover multiple structures contained in signed networks. For this proplem, Yang et al. proposed a signed stochastic block model and presented a variational Bayesian learning method SSL for parameter estimation and model selection \cite{Yang_and_Liu}. 
SSL can effectively discover community or bipartite structure, but it fails to detect coexistence of them.
Zhao et al. proposed a mathematically principled method, namely VBS, to discover the multiple structures in signed networks \cite{Zhao_and_Liu}. 
This method firstly presented a probabilistic model to characterize the signed networks with community, bipartite or coexistence of them, and then deduced the approximate distribution of model parameters by utilizing a variational Bayesian approach. VBS is the classical method of multiple structure discovery for signed networks. 

In recent years, some scholars carried out the researches on multiple structure discovery for unsigned networks. Liu et al. proposed a generative node-attribute network model, namely GNAN, by combining topological information of network and attribute information of nodes\cite{LiuPhysicaA2021}. GNAN can detect communities more accurately due to utilizing node attributes, and detect multiple structures including bipartite, core-periphery, and their mixing structure. 
He et al. developed a Bayesian probabilistic mixture model NEGCD by incorporating network embedding into topological structure of network\cite{HeInformation2021}. NEGCD can detect assortative structure, i.e., community, and disassortative structure, i.e., bipartite, and mixing structure. By contrast, the researches on detecting multiple structures in signed networks are quite few, because it is more difficulty and challenging but worth deeply studying.

Moreover, high time complexity is a common problem faced by all principled methods, making them fail to deal with large-scale networks. To this end, some scholars conducted the researches on scalability. Li et al introduced a new belief-dynamic-based Markov clustering technique, called BMCL, for large-scale network community detection, but BMCL just can only deal with unsigned networks \cite{lihuijia2022}. Zhao et al. proposed a block-wise SBM learning algorithm named as BLOS to improve the scalability of current SBM-based learning methods \cite{ZhaoAAAI15}. 
Different from existing methods, BLOS can implement model selection and parameter estimation simultaneously by introducing the minimum message length (MML) criterion into a block-wise EM algorithm. BLOS has good performance on dealing with large-scale networks in real applications. 
On this basis, Li et al. proposed a reparameterized SBM algorithm RSBM, which is suitable for unsigned networks with heterogeneous distributions of node degree and block size \cite{LiAccess18}.
However, the above methods are merely designed to  focus on structural features of unsigned networks, and can apply to signed networks.

\section{Conclusions}
In this paper, we propose a novel reparameterized signed stochastic block model SSBM to characterize multiple structures in signed networks and present a scalable learning algorithm with model selection ability by integrating MML and CEM. 
The generalization, robustness, and scalability of SSBM are validated on synthetic and real-world networks. 
Experimental results demonstrated the superiority of SSBM by comparing it with five representative network structure discovery methods. 
Future works will be studied from two aspects: the first is studying SBM variations for various networks based on reparameterization, e.g., heterogeneous networks, overlapping networks, multilayer networks, and dynamic networks; 
the second is presenting a general framework of SBM physical parallel learning, which can further improve scalability in dealing with large-scale networks.

%\newpage
\section*{Acknowledgment}
This work was supported in part by the National Natural Science Foundation of China under grants Nos.61902145, 62172185, 61876069, 61976102, and U19A2065; the National Key R\&D Program of China under grants Nos. 2021ZD0112501 and 2021ZD0112502; the International Cooperation Project under grant No. 20220402009GH;  and Jilin Province Natural Science Foundation under Grant No. 20200201036JC.

%\bibliographystyle{cas-model2-names}
%\bibliography{cas-refs}

\end{document}